\documentclass[useAMS,usenatbib]{mn2e}

\usepackage{graphicx}

\title[New members in the Upper Sco association]{
New members in the Upper Sco association from the UKIDSS Early Data Release
\thanks{Based on observations made with the United Kingdom Infrared 
Telescope, operated by the Joint Astronomy Centre on behalf of the 
U.K. Particle Physics and Astronomy Research Council.}
\thanks{Based on observations collected with the ESO 3.6-m/EFOSC2 at 
the European Southern Observatory, La Silla, Chile (ESO programme 
076C-0237)}}
\author[N. Lodieu et al.]{N. Lodieu$^{1}$\thanks{E-mail: nl41@star.le.ac.uk},
N. C. Hambly$^{2}$, \& R. F. Jameson$^{1}$ \\
$^{1}$Department of Physics \& Astronomy, University of Leicester, 
University Road, Leicester LE1 7RH, UK \\ 
$^{2}$Scottish Universities' Physics Alliance (SUPA), 
Institute for Astronomy, School of Physics, University of Edinburgh, \\
~Royal Observatory, Blackford Hill, 
Edinburgh EH9 3HJ, UK  \\
}
\begin{document}

\date{Accepted ---. Received ---; in original form ---}

\pagerange{\pageref{firstpage}--\pageref{lastpage}} \pubyear{2006}

\maketitle

\label{firstpage}

%
%
\begin{abstract}
We present the results of a 9.3 square degree infrared ($ZYJHK$)
survey in the Upper Scorpius association extracted from the UKIRT 
Infrared Deep Sky Survey (UKIDSS) Galactic Cluster Survey
Early Data Release. We have selected a total of 112 candidates
from the ($Z-J$,$Z$) colour-magnitude diagram over the
$Z$ = 12.5--20.5 magnitude range, 
corresponding to M = 0.25--0.01 M$_{\odot}$
at an age of 5 Myr and a distance of 145 pc. Additional photometry
in $J$ and $K$ filters revealed most of them as reddened stars,
leaving 32 possible members. Among them, 15 have proper
motion consistent with higher mass members from Hipparcos
and optical spectra with strong H$\alpha$ in emission and
weak gravity features. We have also extracted two lower mass
candidate members for which no optical spectra are in hand.
Three members exhibit strong H$\alpha$
equivalent widths ($>$20\AA{}), suggesting that they could
still undergo accretion whereas two other dwarfs show signs 
of chromospheric activity.
The likelihood of the binarity of a couple of new stellar
and substellar members is discussed as well.
\end{abstract}


%
%
\section{Introduction}
\label{USco:intro}

The Scorpius Centaurus is the nearest OB association, located 
at a mean distance of 145$\pm$2 pc from the Sun \citep{deBruijne97}.
The association covers 120 square degrees and is composed 
of 3 subgroups: Upper Scorpius, Upper Centaurus Lupus, 
and Lower Centaurus Crux \citep{blaauw64,deZeeuw99}. 
The age of the Upper Sco association is about 5 Myr with little scatter
\citep{preibisch02}. The region is free of extinction with 
Av $\leq$ 2 mag, suggesting that star formation has already ended 
\citep{walter94}.

The first survey conducted in Upper Sco was performed with the 
Einstein X-ray satellite by \citet{walter94}, yielding 28 low-mass
pre-main-sequence stars in 7 square degrees. 
Later on, \citet{deBruijne97} extracted 115 members using 
kinematic information and trigonometric parallaxes for
1215 Hipparcos stars. A larger scale X-ray survey 
carried out by \citet{preibisch98} revealed several hundreds of
X-rays sources confirmed spectroscopically as young
pre-main-sequence members of the Scorpius Centaurus association.
Deeper optical surveys complemented by near-infrared photometry
\citep{ardila00,martin04,slesnick06} crossed the hydrogen-burning
limit with the discovery of members with spectral types later
than M6 \citep{martin96,luhman98}. The current census of
spectroscopic brown dwarfs in Upper Sco includes 46 M6, 10 M7, 
6 M8, and 2 M9 dwarfs, all confirmed as members on the basis 
of their chromospheric activity and weak gravity features
\citep{martin04,slesnick06}.
Recently, we extended the cluster sequence down to 0.008 M$_{\odot}$
using UKIDSS Galactic Cluster Survey science verification data.
A dozen new brown dwarf candidates with spectral types later than
M9 and masses below 0.02 M$_{\odot}$ were extracted over 6.5
square degrees (Lodieu et al., subm.\ to MNRAS).

%
%
%
\begin{figure*}
   \centering
   \includegraphics[width=1.00\linewidth]{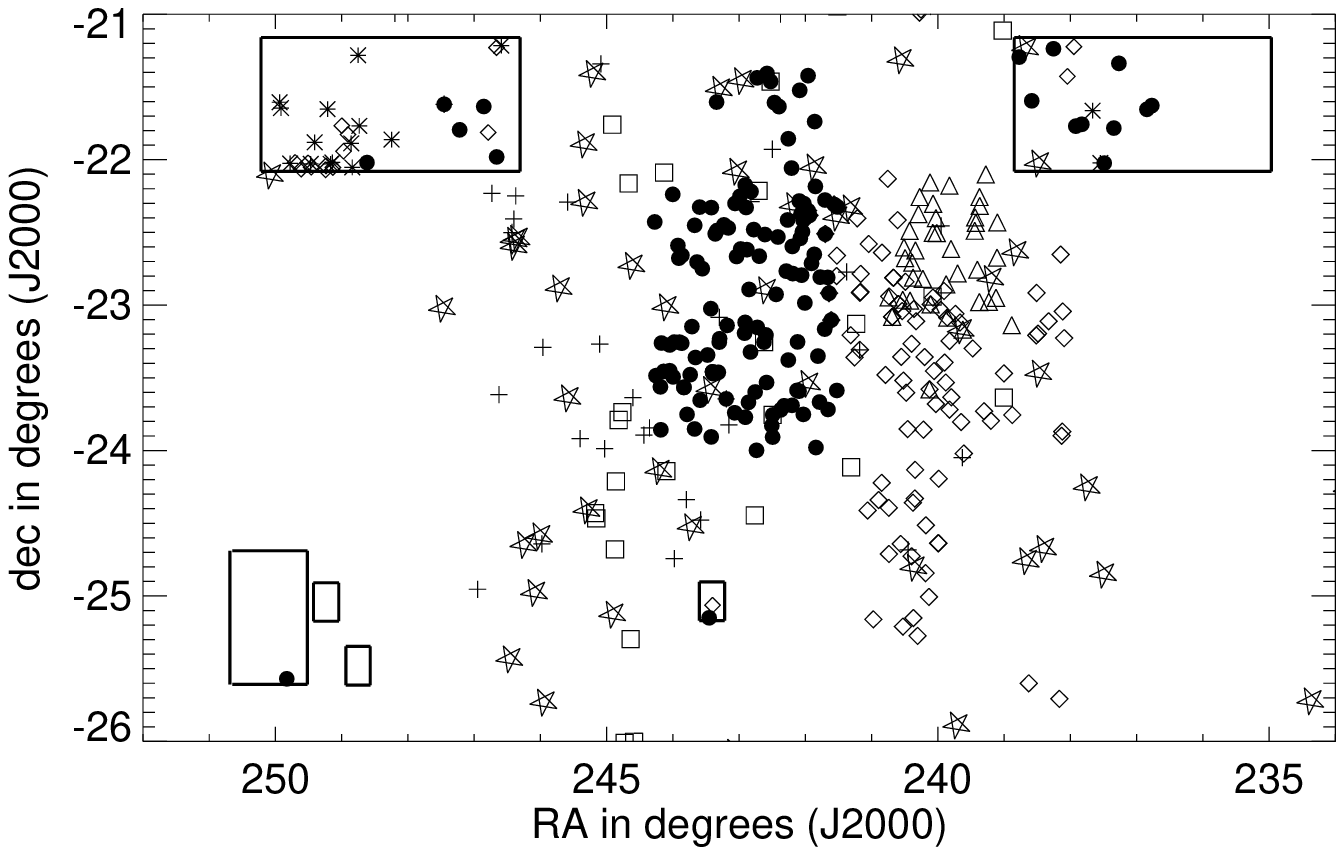}
   \caption{The UKIDSS GCS coverage in the Upper Sco association
            from the Early Data Release is depicted with boxes.
            Early-type stars from Hipparcos are displayed
            as 5-pointed star symbols \citep{deZeeuw99}.
            We have included known members: triangles are from
            \citet{preibisch02}, diamonds from \citet{ardila00},
            squares from \citet{martin04}, and plusses from
            \citet{slesnick06}. The new spectroscopic members
            and non-members from this study are shown as
            filled circles and star symbols, respectively.
            Diamonds located in the area surveyed by the EDR
            are potential reddened members.
            Filled circles outside the boxes represent
            new members of the association extracted from the
            UKIDSS GCS science verification (Lodieu et al.\ 2006).
            Spectroscopic non-members are located in an area affected
            by reddening at (RA,dec)$\sim$(250$^{\circ}$,$-$21.8$^{\circ}$).
            }
   \label{fig_USco:coverage}
\end{figure*}
%

%
%
\begin{table}
 \centering
  \caption{Log of the observations. Central coordinates (in J2000)
           of each WFCAM tile is provided along with the observing
           date. Each tile was observed in all five passbands ($ZYJHK$).
}
 \label{tab_USco:log_obs}
 \begin{tabular}{c c c c c c}
 \hline
Tile  &  R.A.\  & Dec.\ & Date & \\
 \hline
1    &    15    42    &    --21    36    &    2005-06-05     \\
2    &    15    46    &    --21    38    &    2005-06-05     \\
3    &    15    50    &    --21    38    &    2005-06-05     \\
4    &    15    53    &    --21    38    &    2005-06-05     \\
5    &    16    13    &    --25    16    &    2005-06-11     \\
6    &    16    25    &    --25    03    &    2005-06-14    \\
7    &    16    28    &    --21    36    &    2005-06-05     \\
8    &    16    31    &    --21    38    &    2005-06-05     \\
9    &    16    35    &    --21    38    &    2005-06-05     \\
10   &    16    36    &    --25    16    &    2005-06-06     \\
11   &    16    39    &    --21    38    &    2005-06-05     \\
12   &    16    40    &    --25    09    &    2005-06-06    \\
 \hline
 \end{tabular}
\end{table}

The UKIRT Infrared Deep Sky Survey \citep[UKIDSS;][]{lawrence06}
constitutes the new generation of deep large-scale infrared surveys.
The project is subdivided into 5 surveys: The Large Area Survey,
the Galactic Cluster Survey (GCS),
the Galactic Plane Survey, the Deep Extragalactic Survey,
and the Ultra-Deep Survey.
The GCS will survey over thousand square degrees in 10 star-forming
regions and open clusters down to $K$ = 18.4 mag at two epochs.
The main goal of the GCS is to study the Initial Mass Function
well down into the substellar regime in an homogeneous manner
in 10 star-forming regions and open clusters.

In this paper we concentrate on the selection of new cluster member
candidates in a 9.3 square degree area in the Upper
Scorpius association from the UKIDSS Early Data Release 
\citep[EDR;][]{dye06}.
The photometric, proper motion, and spectroscopic results 
presented in this paper
represent the first step towards the ultimate goal of the GCS\@.
In Sect.\ \ref{USco:ukidss_GCS} we briefly present the infrared
photometric observations conducted in Upper Sco.
In Sect.\ \ref{USco:cand} we describe the selection
of new candidates from various colour-magnitude diagrams
complemented by proper motion.
In Sect.\ \ref{USco:spectro} we present the optical spectroscopic
follow-up carried out with the ESO 3.6-m telescope. We infer
spectral types and measure equivalent widths
for the H$\alpha$ emission line and Na{\small{I}} doublet to assess
the membership of the new candidates.
Finally, we summarise our work in Sect.\ \ref{USco:summary}.

%
%
\section{The UKIDSS GCS in Upper Sco}
\label{USco:ukidss_GCS}
%
%
\subsection{The Early Data Release (EDR)}

Twelve WFCAM tiles were observed in the association during the first 
phase of survey observations in June 2005 (Table \ref{tab_USco:log_obs}),
covering a total of 9.3 square degrees
(Fig.\ \ref{fig_USco:coverage}) in $ZYJHK$ broad-band filters
\citep{hewett06}.
A summary of the observations, including the 12 WFCAM tiles, central 
coordinates and date of observations
is provided in Table \ref{tab_USco:log_obs}. Uniformity of quality in
the UKIDSS EDR (e.g.\ seeing, photometricity, etc.) is discussed at
length in Dye et al.\ (2006).
We refer the reader to \citet{lawrence06} for more details on the 
observing strategy of the UKIDSS\footnote{The main page is www.ukidss.org} 
project and its individual components.

All observations carried out within the framework of the UKIDSS project
are pipeline-processed at the Cambridge Astronomical Survey Unit
(CASU; Irwin et al.\ 2006, in prep.)\footnote{The CASU WFCAM 
webpage can be found at http://apm15.ast.cam.ac.uk/wfcam}.
The processed data are then archived in Edinburgh and released to the
European community through the WFCAM Science Archive
(WSA; Hambly et al.\ 2006, in prep.)\footnote{The WFCAM 
Science Archive is accessible at http://surveys.roe.ac.uk/wsa}.

%
%
\subsection{Observations and colour-magnitude diagrams}

The UKIDSS GCS observations taken in Upper Sco and included 
in the Early Data Release amounts for a total 
of 9.3 square degrees (Fig.\ \ref{fig_USco:coverage}).
There are 3 main blocks centered approximately at 
(RA,dec)=(237.0,$-$21.55), (248.0,$-$21.55), and
(250.05,$-$25.15) with three additional small blocks 
(made of 2 pawprints) contributing for 0.3 deg$^{2}$ in total.
The survey was conducted in $ZYJHK$ and reached 5$\sigma$
completeness limits of $Z$ = 20.0, $Y$ = 19.6,
$J$ = 18.6, $H$ = 18.1, and $K$ = 17.5 mag. Hence, this survey
is 3 magnitudes deeper than any previous study in the region
and is able to probe young L dwarfs according to the DUSTY 5 Myr
isochrones.
The two lowest mass spectroscopic members reported in Upper Sco 
to date have spectral types of M9 corrsponding to
$\sim$20 Jupiter masses \citep{martin04,slesnick06}.

\begin{figure*}
   \centering
   \includegraphics[width=0.49\linewidth]{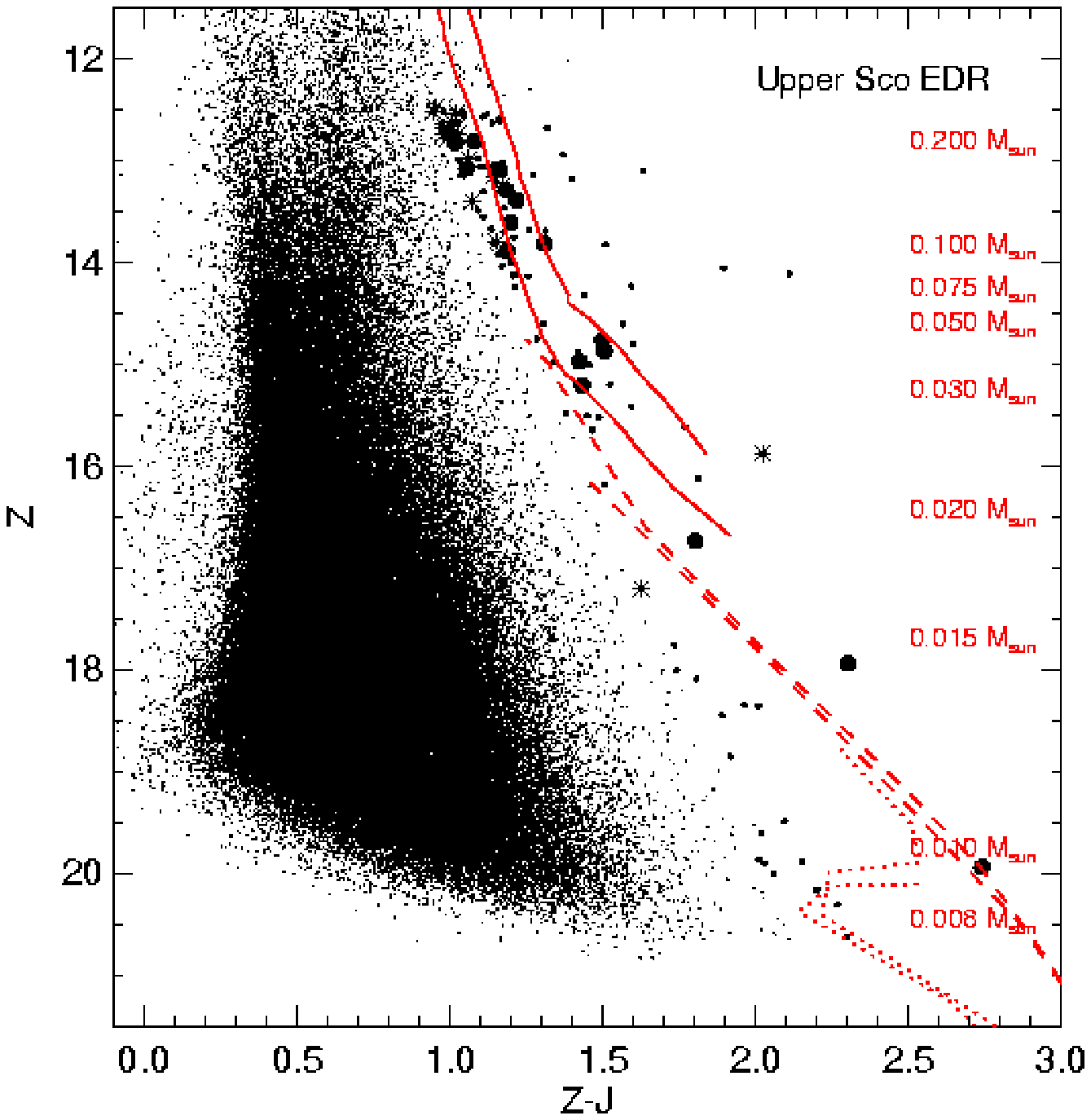}
   \includegraphics[width=0.49\linewidth]{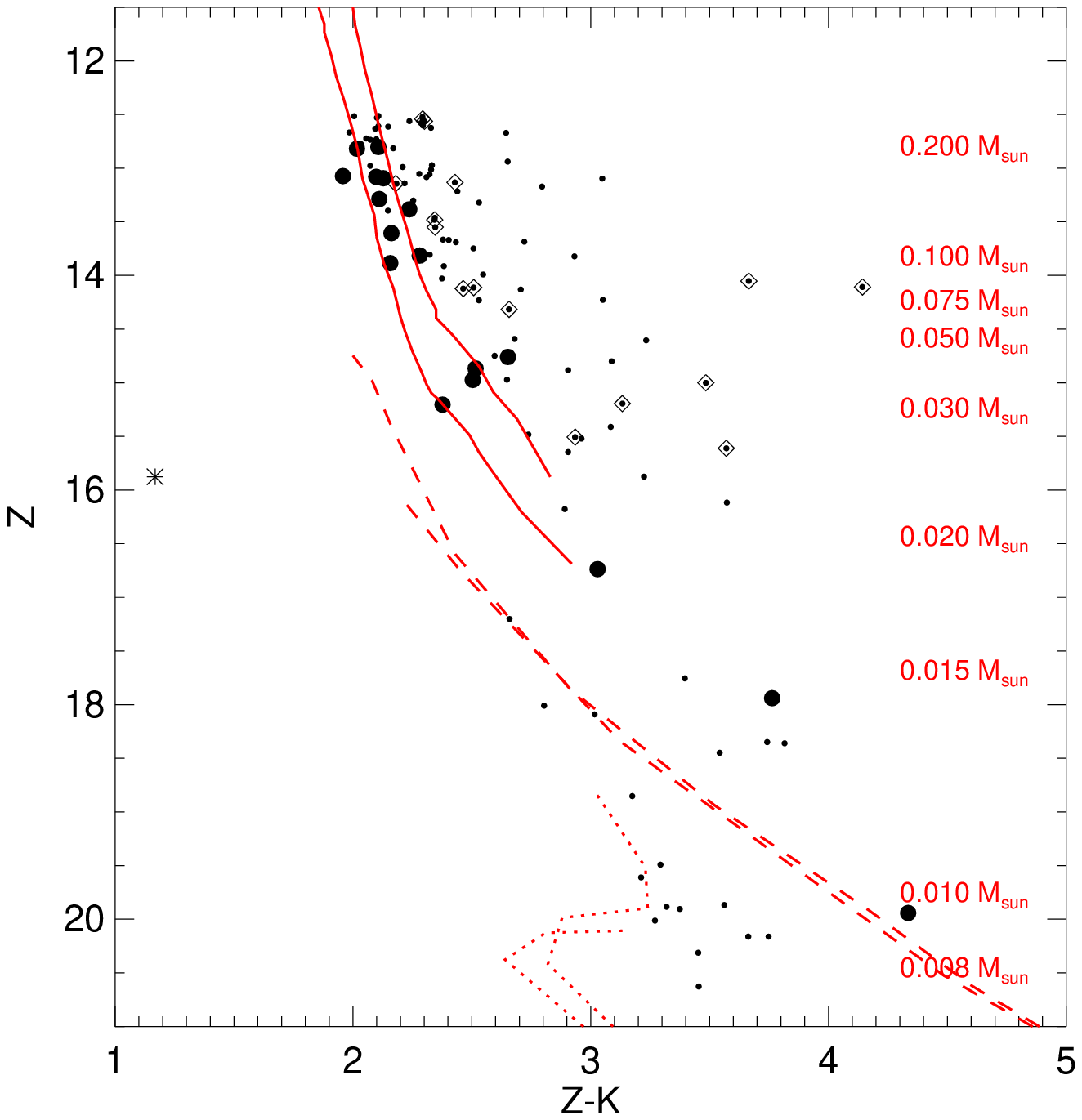}
   \includegraphics[width=0.49\linewidth]{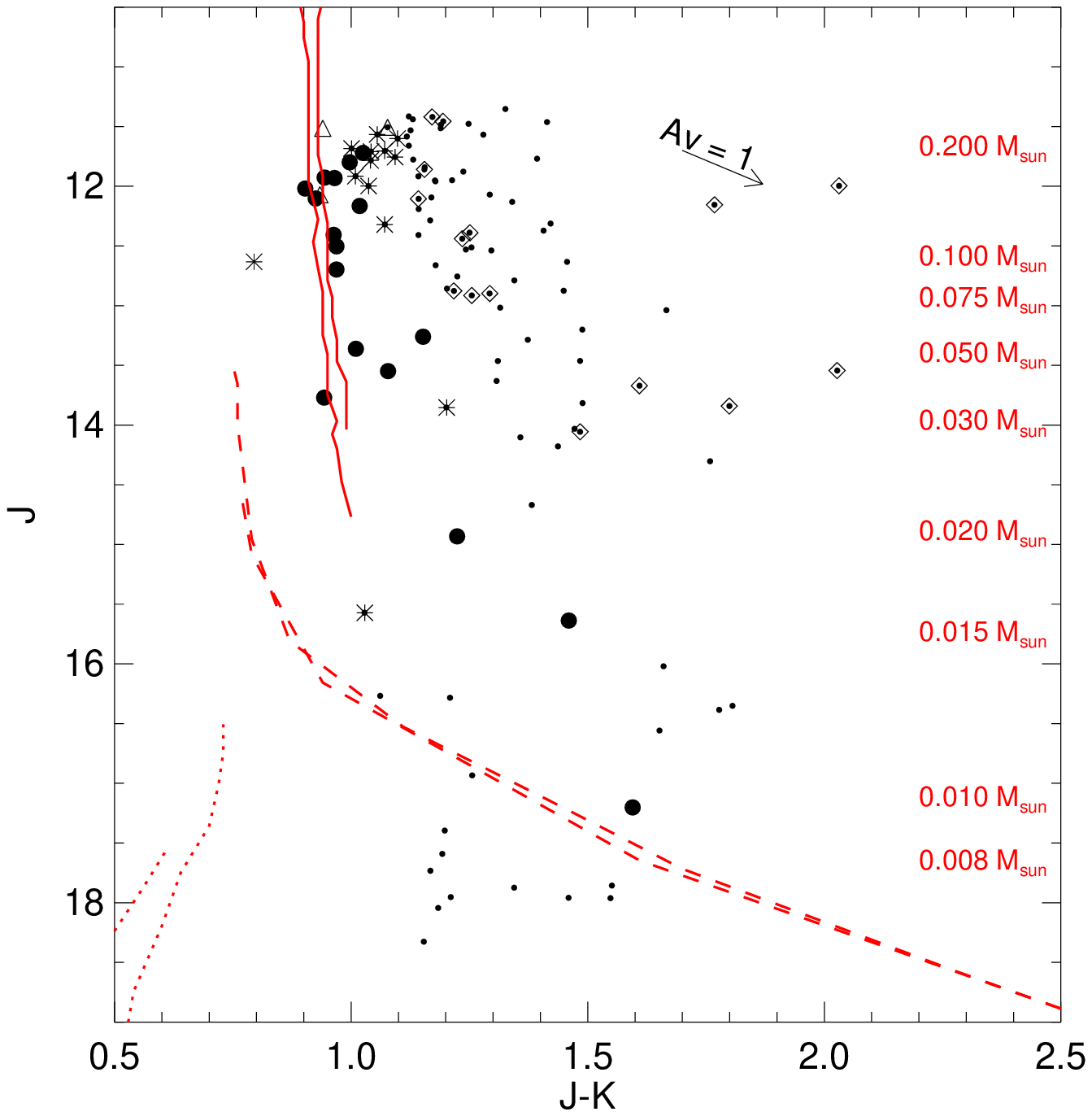}
   \includegraphics[width=0.49\linewidth]{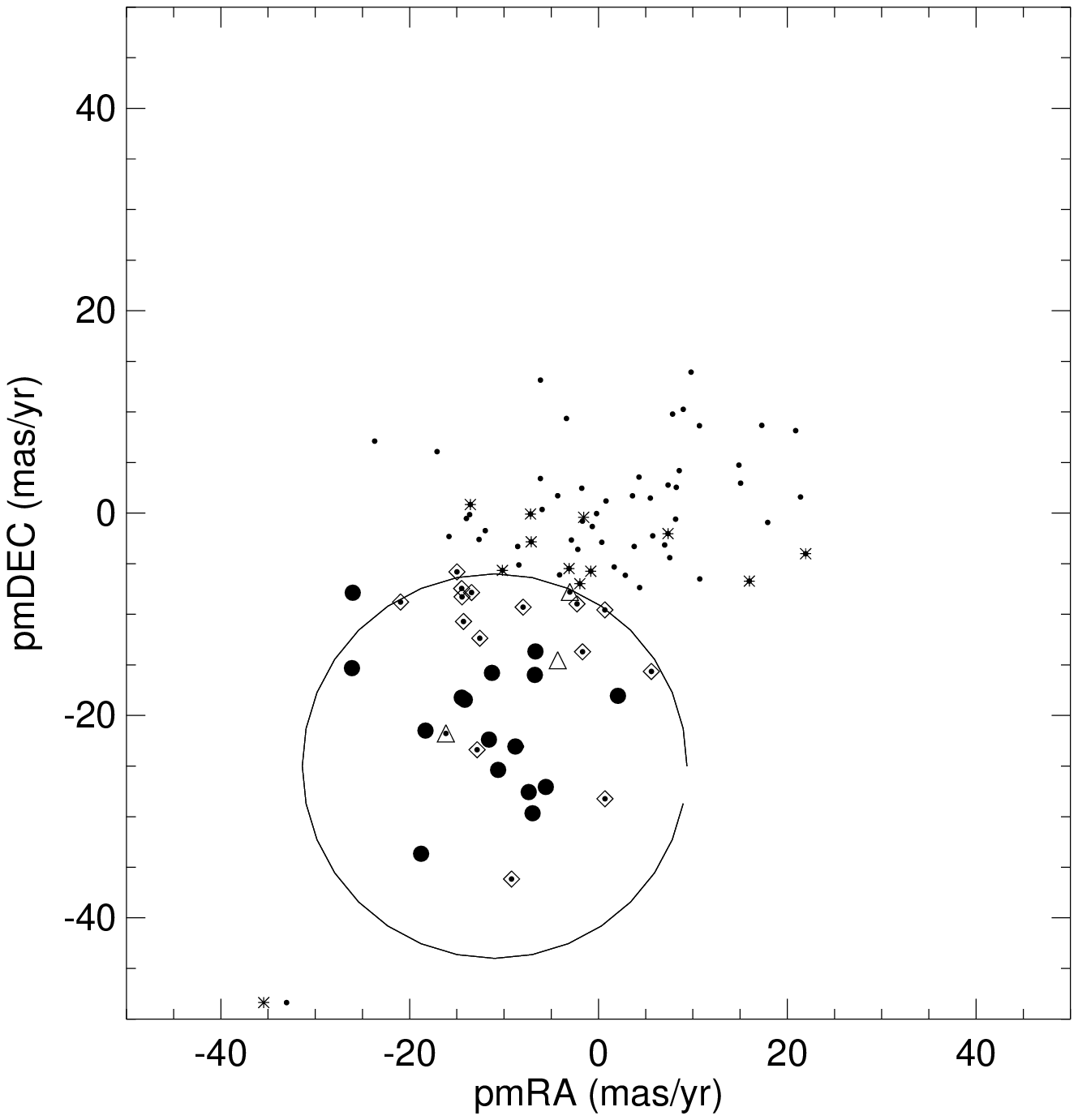}
   \caption{Colour-magnitude diagrams for 9.3 square degrees 
   in the Upper Scorpius assocation from the UKIDSS Early Data Release.
   Large filled circles are new spectroscopic Upper Sco members. 
   Open triangles and star symbols are spectroscopic and
   proper motion non-members, respectively. The dots represent
   all candidates classified photometrically as reddened stars
   from their location in the ($J-K$,$J$) and ($Z-K,J-K$) diagrams.
   The diamonds are possible reddened members for which no
   optical spectroscopy is currently available.
   Overplotted are 
   the 5 and 10 Myr NextGen \citep[solid line;][]{baraffe98},
   DUSTY \citep[dashed line;][]{chabrier00c}, and
   COND \citep[dotted line;][]{baraffe02} isochrones.
   The mass scale is shown on the right hand side of the diagrams
   and spans 0.3--0.008 M$_{\odot}$.
   {\it{Top left:}} ($Z-J$,$Z$) diagram.
   {\it{Top right:}} ($Z-K$,$Z$) diagram.
   {\it{Bottom left:}} ($J-K$,$J$) diagram.
   {\it{Bottom right:}} Vector point diagram for sources brighter
   than $J$ = 15.8 mag because the faintest ones do not have proper 
   motion measurements due to the lack of 2MASS detection.
   The last three diagrams contain only photometric candidates
   selected in the ($Z-J$,$Z$) diagram.
}
   \label{fig_USco:cmds}
\end{figure*}

We have selected from the EDR point sources (Class parameter set to
$-$1 or $-$2 in all passbands) detected in all 5 filters ($ZYJHK$)
by use of the SQL script given in Appendix~\ref{sql} injected 
in the WFCAM Science Archive.
The total number of point sources is 174,010 (dots in the
colour-magnitude diagrams in Fig.\ \ref{fig_USco:cmds})
in the $Z$ = 11.4--20.5 magnitude range, corresponding to masses
between 0.35 M$_{\odot}$ and 0.008 M$_{\odot}$ at an age of
5 Myr and a distance of 145 pc \citep{baraffe98,chabrier00c}.

The colour-magnitude diagrams resulting from the GCS observations
in Upper Sco are shown in Fig.\ \ref{fig_USco:cmds}, including the
($Z-J$,$Z$), ($Z-K$,$Z$), and ($J-K$,$J$) diagrams.
Overplotted are 5 and 10 Myr NextGen \citep[solid lines;][]{baraffe98},
DUSTY \citep[dashed lines;][]{chabrier00c}, and
COND \citep[dotted lines;][]{baraffe02} isochrones shifted
at the distance of the association (d = 145 pc). Note that isochrones
were computed for the UKIDSS passbands (courtesy I.\ Baraffe and
F.\ Allard).

%
%
\section{New candidate members of the Upper Sco association}
\label{USco:cand}

This section describes the selection of new candidates in the
Upper Scorpius association using colour-magnitude diagrams and 
proper motions.

%
%
%

%
%
\subsection{Selection of candidates in Upper Sco}
\label{USco:select}
Candidates in open clusters are generally selected to the
right of the theoretical isochrones or the
Zero-Age-Main-Sequence \citep{leggett92}.
To select all possible members in Upper Sco at the
expense of including some contaminants, we have extracted all 
candidates to the right of a straight line running from
($Z-J$,$Z$)=(0.8,11.5) to (2.1,21.5). The total
number of candidates is 192\@. However, we have limited
our analysis to objects fainter than $Z$ = 12.5
($J \sim$ 11.5), leaving 112 candidates. This threshold
translates into a mass of 0.25 M$_{\odot}$ and corresponds
to  completeness limit of the survey conducted by \citet{preibisch02}
in the region. We provide an electronic table with all photometric
candidates, including those classified later as non-members after 
proper motion and spectroscopic analysis (Table \ref{tab_USco:cand_phot}).
Some candidates fainter than $Z$ = 17 mag
are running along the straight line defined as our
criterion and located to the left of the theoretical
isochrones \citep{chabrier00c} in the ($Z-J$,$Z$) diagram. 
Consequently, we cast doubt
on the membership of those objects and do not consider
them in the subsequent analysis for two reasons:
first, one candidate located at $Z \sim$ 16.2 and
$Z-J \sim$ 1.5 turned out as a non-member after optical 
spectroscopy, and, second, the cluster sequence runs
further to the red and to the right of the isochrones as
demonstrated in 
Lodieu et al.\ (2006, subm.\ to MNRAS).

To further assess the membership of the 112 candidates,
we have investigated their location in the ($Z-K$,$Z$)
and ($J-K$,$J$) colour-magnitude diagrams (top right
and bottom left graphs in Fig.\ \ref{fig_USco:cmds},
respectively) as well as in the ($Z-K$,$J-K$) two-colour
diagram. The ($Z-K$,$Z$) diagram is characterised
by a spread of the sample of candidates which can be
explained by the presence of interstellar reddening
in the part of Upper Sco released in the Early Data Release.
This trend is confirmed in the ($J-K$,$J$) diagram
where we indicated a reddening vector of A$_{V}$ = 1 and
by the location of those sources in Fig. \ref{fig_USco:coverage}.
As a consequence, we have applied the following colour 
criteria to weed out reddened sources:
\begin{itemize}
\item 0.90 $\leq J-K \leq$ 1.10 for $J$ = 11.5--13.0 $\Rightarrow$ 23 candidates
\item 0.90 $\leq J-K \leq$ 1.25 for $J$ = 13.0--14.0 $\Rightarrow$ 5 candidates
\item All 4 sources fainter than $J$ = 14 mag were kept
\end{itemize}
A total of 32 sources remain as photometric candidates. 
Their membership is further investigated using proper
motions (Sect.\ \ref{USco:PMs}) and optical spectroscopy
(Sect.\ \ref{USco:spectro}) as additional criteria.

%
%
\begin{table*}
 \centering
  \caption{All 112 photometric candidates selected from the
($Z-J$,$Z$) colour-magnitude diagram in 9.3 square degrees
in the Upper Sco association. This table provides coordinates
(J2000), $ZYJHK$ photometry from the GCS, and proper motion
from the GCS/2MASS cross-correlation for members and non-members.
The table is available in electronic format only.
}
 \label{tab_USco:cand_phot}
 \begin{tabular}{c c c c c c c c c}
 \hline
R.A. & Dec  &  $Z$  &  $Y$  &  $J$  &  $H$  & $K$ & $\mu_{\alpha}\cos{\delta}$ & $\mu_\delta$ \\
 \hline
16:26:38.63 & $-$21:58:51.7 & 12.803  & 12.246  & 11.721  & 11.061  & 10.696  & $-$18.32 & $-$21.49 \cr
\dots{}     & \dots{}       & \dots{} & \dots{} & \dots{} & \dots{} & \dots{} & \dots{}  & \dots{}  \cr
\dots{}     & \dots{}       & \dots{} & \dots{} & \dots{} & \dots{} & \dots{} & \dots{}  & \dots{}  \cr
16:28:52.80 & $-$21:47:41.1 & 12.819  & 12.266  & 11.799  & 11.109  & 10.802  & $-$26.03 &  $-$7.87 \cr
 \hline
 \end{tabular}
\end{table*}
%

%
%
%
\begin{table*}
 \centering
  \caption{Seventeen new members of the Upper Sco association 
confirmed from their photometry, proper motion, and spectroscopy
(filled circles in Fig.\ \ref{fig_USco:cmds}). 
This table lists the name of the object according to the IAU nomenclature, 
the equatorial coordinates (J2000), the magnitudes from the GCS,
the proper motions (in mas/yr), the H$\alpha$ and Na{\small{I}} 
equivalent widths (in \AA{}), 
and the spectral types with an accuracy of half a subclass.
The two faintest candidates ($Z >$ 17.0 mag) have neither spectral types 
nor equivalent width measurements because they are too faint for 
optical spectroscopy with ESO 3.6-m/EFOSC2 but one 
has a proper motion consistent with the association.
}
 \label{tab_USco:Members}
 \begin{tabular}{@{\hspace{1mm}}l c c c c c c c c c c c c@{\hspace{1mm}}}
 \hline
USco J\ldots{} & R.A. & Dec  &  $Z$  &  $Y$  &  $J$  &  $H$  & $K$ & $\mu_{\alpha}\cos{\delta}$ & $\mu_\delta$ & H$\alpha$ & Na{\small{I}} & SpT \\
 \hline
1626$-$2158 & 16:26:38.63 & $-$21:58:51.7 & 12.803 & 12.246 & 11.721 & 11.061 & 10.696 & $-$18.32 & $-$21.49 &  $-$4.70 &  2.56 & M4.0 \cr
1628$-$2147 & 16:28:52.80 & $-$21:47:41.1 & 12.819 & 12.266 & 11.799 & 11.109 & 10.802 & $-$26.03 &  $-$7.87 &  $-$4.40 &  3.55 & M3.5 \cr
1549$-$2146 & 15:49:22.59 & $-$21:46:57.6 & 13.074 & 12.536 & 12.019 & 11.387 & 11.116 & $-$26.11 & $-$15.32 &  $-$9.10 &  4.40 & M3.5 \cr
1553$-$2114 & 15:53:01.32 & $-$21:14:13.7 & 13.081 & 12.486 & 11.927 & 11.329 & 10.983 & $-$10.63 & $-$25.37 & $-$30.70 &  2.07 & M4.0 \cr
1551$-$2146 & 15:51:40.32 & $-$21:46:10.6 & 13.094 & 12.476 & 11.931 & 11.382 & 10.966 &  $-$7.39 & $-$27.57 &  $-$9.00 &  2.41 & M4.0 \cr
1551$-$2145 & 15:51:18.70 & $-$21:45:23.7 & 13.288 & 12.651 & 12.102 & 11.472 & 11.177 & $-$14.48 & $-$18.23 &  $-$9.60 &  3.12 & M4.0 \cr
1627$-$2138 & 16:27:25.52 & $-$21:38:03.8 & 13.384 & 12.751 & 12.165 & 11.531 & 11.147 & $-$11.60 & $-$22.38 &  $-$5.90 &  3.38 & M4.0 \cr
1629$-$2137 & 16:29:48.78 & $-$21:37:08.9 & 13.606 & 12.957 & 12.407 & 11.845 & 11.444 & $-$18.79 & $-$33.66 & $-$30.00 &  3.25 & M5.0 \cr
1555$-$2117 & 15:55:05.31 & $-$21:17:40.4 & 13.815 & 13.136 & 12.503 & 11.900 & 11.534 & $-$14.15 & $-$18.44 & $-$13.00 &  2.42 & M5.0 \cr
1547$-$2137 & 15:47:04.94 & $-$21:37:40.5 & 13.885 & 13.230 & 12.697 & 12.051 & 11.728 &     2.06 & $-$18.06 & $-$10.80 &  3.78 & M4.0 \cr
1634$-$2201 & 16:34:28.50 & $-$22:01:12.1 & 14.760 & 13.953 & 13.260 & 12.612 & 12.108 &  $-$8.80 & $-$23.06 & $-$13.60 &  3.70 & M6.0 \cr
1549$-$2201 & 15:49:57.33 & $-$22:01:25.7 & 14.867 & 13.989 & 13.361 & 12.792 & 12.351 &  $-$6.68 & $-$13.67 & $-$12.30 &  2.88 & M6.0 \cr
1613$-$2509 & 16:13:48.80 & $-$25:09:00.9 & 14.974 & 14.202 & 13.548 & 12.999 & 12.470 &  $-$6.98 & $-$29.66 & $-$25.50 &  2.37 & M5.0 \cr
1549$-$2120 & 15:49:04.14 & $-$21:20:15.2 & 15.204 & 14.432 & 13.770 & 13.212 & 12.827 & $-$11.28 & $-$15.79 & $-$23.00 &  3.20 & M6.0 \cr
1554$-$2135 & 15:54:19.99 & $-$21:35:43.1 & 16.737 & 15.726 & 14.932 & 14.282 & 13.708 &  $-$5.58 & $-$27.06 & $-$29.40 &  2.80 & M8.0 \cr
1547$-$2139 & 15:47:22.82 & $-$21:39:14.3 & 17.940 & 16.559 & 15.637 & 14.834 & 14.177 &  $-$6.74 & $-$15.99 &    ---   &  ---  &  --- \cr
1639$-$2534 & 16:39:19.15 & $-$25:34:09.9 & 19.942 & 18.328 & 17.202 & 16.395 & 15.607 &  ---     &   ---    &    ---   &  ---  &  --- \cr
 \hline
 \end{tabular}
\end{table*}

%
%
%
\begin{table*}
 \centering
  \caption{This table lists the equatorial coordinates
(in J2000), the $ZYJHK$ magnitudes, and the proper motion
of 15 photometric candidates classified as non-members after
optical spectroscopic follow-up (star symbols and open triangles
in Fig.\ \ref{fig_USco:cmds}). The spectra look like
reddened early-type stars apart from two objects classified
as a M5 dwarf (15 50 39.55, $-$21 39 47.5) and a late-M dwarf
(15 50 11.5, $-$22 01 21.9).
All spectroscopic non members but 3 exhibit
proper motions discrepant with the mean motion of the
association.
}
 \label{tab_USco:NM}
 \begin{tabular}{c c c c c c c c c}
 \hline
R.A.\ & Dec.\  &  $Z$  &  $Y$  &  $J$  &  $H$  & $K$ & $\mu_{\alpha}\cos{\delta}$ & $\mu_\delta$ \\ \hline
16:37:38.13 & $-$21:52:50.2 & 12.516 & 12.097 & 11.566 & 10.838 & 10.511 & $-$10.17 &  $-$5.66 \cr
16:37:52.30 & $-$22:01:43.9 & 12.530 & 12.054 & 11.506 & 10.793 & 10.429 &  $-$3.03 &  $-$7.79 \cr
16:35:21.87 & $-$22:03:12.0 & 12.609 & 12.139 & 11.601 & 10.798 & 10.503 &  $-$7.20 &  $-$0.09 \cr
16:39:45.13 & $-$21:36:19.0 & 12.667 & 12.202 & 11.683 & 11.004 & 10.682 &  21.95 &  $-$4.01 \cr
16:39:06.86 & $-$22:01:33.8 & 12.723 & 12.233 & 11.710 & 10.967 & 10.668 &  $-$0.81 &  $-$5.75 \cr
16:34:56.41 & $-$21:46:04.0 & 12.731 & 12.255 & 11.704 & 10.926 & 10.633 &  $-$1.98 &  $-$6.97 \cr
16:35:25.71 & $-$21:53:15.4 & 12.735 & 12.253 & 11.755 & 11.041 & 10.662 &  $-$3.11 &  $-$5.48 \cr
16:39:40.12 & $-$21:38:42.6 & 12.775 & 12.304 & 11.783 & 11.025 & 10.742 &  15.98 &  $-$6.72 \cr
16:36:33.90 & $-$22:01:08.4 & 12.979 & 12.453 & 11.915 & 11.237 & 10.906 &  $-$7.13 &  $-$2.83 \cr
16:36:40.01 & $-$22:01:46.4 & 13.145 & 12.571 & 11.998 & 11.291 & 10.961 &   7.36 &  $-$2.03 \cr
15:50:39.55 & $-$21:39:47.5 & 13.245 & 12.587 & 12.071 & 11.465 & 11.137 & $-$16.16 & $-$21.77 \cr
16:32:59.62 & $-$21:51:41.9 & 13.396 & 12.877 & 12.320 & 11.511 & 11.249 & $-$13.57 &   0.85 \cr
16:36:51.29 & $-$21:39:08.3 & 13.786 & 13.111 & 12.632 & 12.026 & 11.837 &  $-$1.56 &  $-$0.41 \cr
16:35:00.85 & $-$21:16:55.3 & 15.876 & 14.708 & 13.853 & 13.273 & 12.651 & $-$35.45 & $-$48.37 \cr
15:50:11.50 & $-$22:01:21.9 & 17.202 & 16.283 & 15.572 & 14.969 & 14.543 & $-$20.47 & $-$80.92 \cr
 \hline
 \end{tabular}
\end{table*}

%
%
%
\begin{table*}
 \centering
  \caption{This table lists the equatorial coordinates
(in J2000), the $ZYJHK$ magnitudes, and the proper motion
of 15 potential reddened members in Upper Sco whose proper
motion is consistent with the association (diamonds
in Fig.\ \ref{fig_USco:cmds}).
}
 \label{tab_USco:reddened_cand}
 \begin{tabular}{c c c c c c c c c}
 \hline
R.A.\ & Dec.\  &  $Z$  &  $Y$  &  $J$  &  $H$  & $K$ & $\mu_{\alpha}\cos{\delta}$ & $\mu_\delta$ \\ \hline
16:38:26.94 & $-$22:03:55.8 &  12.541 & 11.997 & 11.419 & 10.680 & 10.248 &    $-$1.687 &   $-$13.701 \cr 
16:13:36.51 & $-$25:03:47.4 &  12.563 & 12.056 & 11.456 & 10.614 & 10.262 &    $+$5.603 &   $-$15.645 \cr 
16:26:39.26 & $-$21:13:45.4 &  13.133 & 12.451 & 11.859 & 11.133 & 10.704 &   $-$14.451 &    $-$8.276 \cr 
16:37:50.68 & $-$22:03:07.9 &  13.143 & 12.612 & 12.105 & 11.292 & 10.963 &    $-$7.985 &    $-$9.292 \cr 
16:38:01.92 & $-$22:01:18.9 &  13.482 & 12.970 & 12.389 & 11.461 & 11.138 &    $+$0.675 &    $-$9.565 \cr 
16:36:56.95 & $-$22:02:32.1 &  13.550 & 13.026 & 12.439 & 11.541 & 11.204 &   $-$14.490 &    $-$7.445 \cr 
16:35:53.43 & $-$21:56:31.2 &  14.052 & 13.138 & 12.155 & 11.046 & 10.387 &    $-$2.271 &    $-$8.968 \cr 
16:36:32.01 & $-$22:03:13.0 &  14.108 & 13.070 & 11.996 & 10.644 &  9.965 &   $-$14.992 &    $-$5.803 \cr 
16:36:58.40 & $-$22:04:12.1 &  14.113 & 13.505 & 12.898 & 11.912 & 11.605 &   $-$20.965 &    $-$8.786 \cr 
16:35:37.41 & $-$21:49:36.3 &  14.123 & 13.519 & 12.914 & 12.019 & 11.659 &   $-$12.573 &   $-$12.365 \cr 
16:27:09.41 & $-$21:48:45.7 &  14.316 & 13.529 & 12.876 & 12.113 & 11.659 &   $-$13.443 &    $-$7.843 \cr 
15:51:47.09 & $-$21:13:23.6 &  15.000 & 14.310 & 13.543 & 12.440 & 11.516 &   $-$12.856 &   $-$23.383 \cr 
16:35:59.71 & $-$21:45:59.0 &  15.194 & 14.441 & 13.670 & 12.484 & 12.061 &    $+$0.691 &   $-$28.230 \cr 
16:38:48.92 & $-$22:01:07.0 &  15.506 & 14.793 & 14.056 & 12.965 & 12.572 &   $-$14.298 &   $-$10.709 \cr 
15:52:10.88 & $-$21:25:37.5 &  15.611 & 14.729 & 13.840 & 13.068 & 12.041 &    $-$9.209 &   $-$36.159 \cr
 \hline
 \end{tabular}
\end{table*}
%

%
%
\subsection{Proper motions}
\label{USco:PMs}
Optical or near-infrared photometry alone is not sufficient 
to extract a clean sample of members in open clusters. 
We have exploited the 2MASS database and the GCS observations 
as first and second epoch, respectively, to compute proper motions
for candidates brighter than $J$ = 15.8 mag, corresponding to 
the 2MASS 5$\sigma$ completeness limit and a mass
of 0.015 M$_{\odot}$ in Upper Sco \citep{chabrier00c}. 
The vector point diagram (proper motion in right
ascension versus proper motion in declination)
is presented in the bottom right panel in
Fig.\ \ref{fig_USco:cmds}.
Most of the sources classified as reddened stars exhibit 
a small proper motion (small dots), hence confirming 
their status of distant objects and non-members.
However, there remain 15 objects 
(diamonds; Table \ref{tab_USco:reddened_cand})
considered as reddened stars
in the previous section which do have a proper motion within
the 2$\sigma$ circle plotted in the vector point diagram
in Fig.\ \ref{fig_USco:cmds}. Although most of them are
likely to be reddened stars because they lie in the same
part of the diagram as the spectroscopic members
(Fig.\ \ref{fig_USco:coverage}), they might
be some reddened members. Unfortunately we do not provide
optical spectroscopy for them but they need to be
followed-up to determine how complete is the sample selected 
from the sequence alone.
Two other groups of sources for which we obtained
optical spectra (see Sect.\ \ref{USco:spectro} for more
details) are overplotted. On the one hand,
12 candidates classified as non-members and plotted as 
star symbols in Fig.\ \ref{fig_USco:cmds} have small proper motion 
close to ($\mu_{\alpha}\cos{\delta}$,$\mu_{\delta}$) = (0,0) mas/yr. 
On the other hand, 20 new candidates shown as filled circles 
and triangles in Fig.\ \ref{fig_USco:cmds} exhibit
proper motion consistent with the mean motion of the
association from Hipparcos \citep{deBruijne97,preibisch98}.
The faintest photometric member candidate is too faint
to compute its proper motion from the 2MASS images.

%
%
\section{Spectroscopic follow-up}
\label{USco:spectro}

This section describes the spectroscopic observations
of new candidates in Upper Sco and discuss their membership.

%
%
\subsection{Spectroscopic observations}
\label{USco:spectro_obs}

We have carried out low-resolution optical spectroscopy
of 31 candidates in Upper Sco with EFOSC2
on the ESO 3.6-m telescope in La Silla, Chile. The observations 
were obtained over 5 nights on 27 March-01 April 2006\@.
Conditions were photometric during the first three nights whereas
the last two nights were affected by thin cirrus. Seeing was typical
in the 0.5--1.0 arcsec range during the observing run.

The EFOSC2 detector is a 2048\,$\times$\,2048 pixel MIT CCDs with
a pixel size of 0.157$''$/pix, yielding a 5.5\,$\times$\,5.5 arcmin 
field-of-view. We have employed the grism \#16 with a 1 arcsec slit
to cover the wavelength range from 6000 to 10300 \AA{} at a 
resolution of R$\sim$300\@.  
We have obtained an internal flat-field after each 
spectrum to remove fringing longwards of 7500 \AA{}.
Each target was observed once with exposure 
times ranging from 240\,sec for the brightest
objects to 3600\,sec for the faintest objects, respectively.
In addition, we have observed 3 known Upper Sco members:
SCH160147$-$244101 \citep[M5;][]{slesnick06},
DENIS J160514$-$240653 \citep[M6;][]{martin04},
and SCH162528$-$165851 \citep[M8;][]{slesnick06} as well as
a M5 field dwarf (SCH160311$-$134544)
classified as a non-member by \citet{slesnick06}.
A couple of other field dwarfs were observed, including
Gl643 (M3.5), Gl699 (M4), Gl166 (M4.5), GJ406 (M6),
and vB10 (M8). Spectrophotometric standard stars 
\citep[HR4963, LTT3218, LTT3864, and EG274;][]{hamuy92}
were observed during the night with slits of 1 and 5 arcsec to 
calibrate our targets. Bias, dome flats, and arc lamps were 
observed during the afternoon before each night. 

%
%
\subsection{Data reduction}
\label{USco:spectro_reduc}

The data reduction was carried out under the IRAF environment
following standard procedures.
First, we combined 10 bias frames and subtracted the averaged bias from
the combined flat field and the science target. Second, we have divided 
the science frame by the normalised response function of the
internal flat field observed immediately after each target.
Afterwards, we extracted a one-dimensional spectrum interactively 
with apsum by choosing the appropriate aperture and background 
intervals. Then, we used the arc lamps to calibrate our spectra,
yielding typical rms better than 0.35\AA{}.
The flux calibration with the spectrophotometric standard stars was
not possible due to the presence of 2$^{\rm nd}$ order contamination
longwards of 7200 \AA{}. All four standard targeted during the observing
are white dwarfs or early-type stars, yielding an overestimate of the
detector response in the red part of the spectrum. To solve this issue,
we used the optical spectra of 3 Upper Sco members\footnote{Spectra
kindly provided by Eduardo Mart{\'{\i}}n and Catherine Slesnick}
to correct the shape of our spectra. However, the amount of flux
from the blue end of the M dwarf spectra kicks in redwards
of $\sim$9000\AA{}, making the calibration very difficult.
Thus, the shape of the science target and then comparison
with templates is only valid between 6400\AA{} and 8500\AA{}.
No correction for telluric bands was made to the spectra.
The optical spectra (normalised at 7500\AA{}) of Upper Sco 
members are displayed in Fig.\ \ref{fig_USco:spectra}.

%
%
\subsection{Spectral classification}
\label{USco:spectro_SpT}

The classification of M dwarfs is generally based on spectral indices 
measuring the strength of molecular absorption bands and atomic lines
\citep{martin99a,kirkpatrick99}. Spectral indices are defined over a 
large wavelength range and are valid for old field dwarfs.
The effect of gravity can influence the computation of
the indices \citep{martin96}.
Thus, we preferred to rely on the direct comparison with 
known Upper Sco members and field M dwarfs observed with 
the same telescope/instrument set-up in order to classify
our targets. We used 3 known Upper Sco members, 
SCH1601$-$2441 \citep[M5;][]{slesnick06},
DENIS1605$-$2406 \citep[M6;][]{martin04}, and
SCH1625$-$1658 \citep[M8;][]{slesnick06}, as young M
dwarf templates.

\begin{figure*}
   \centering
   \includegraphics[width=1.0\linewidth]{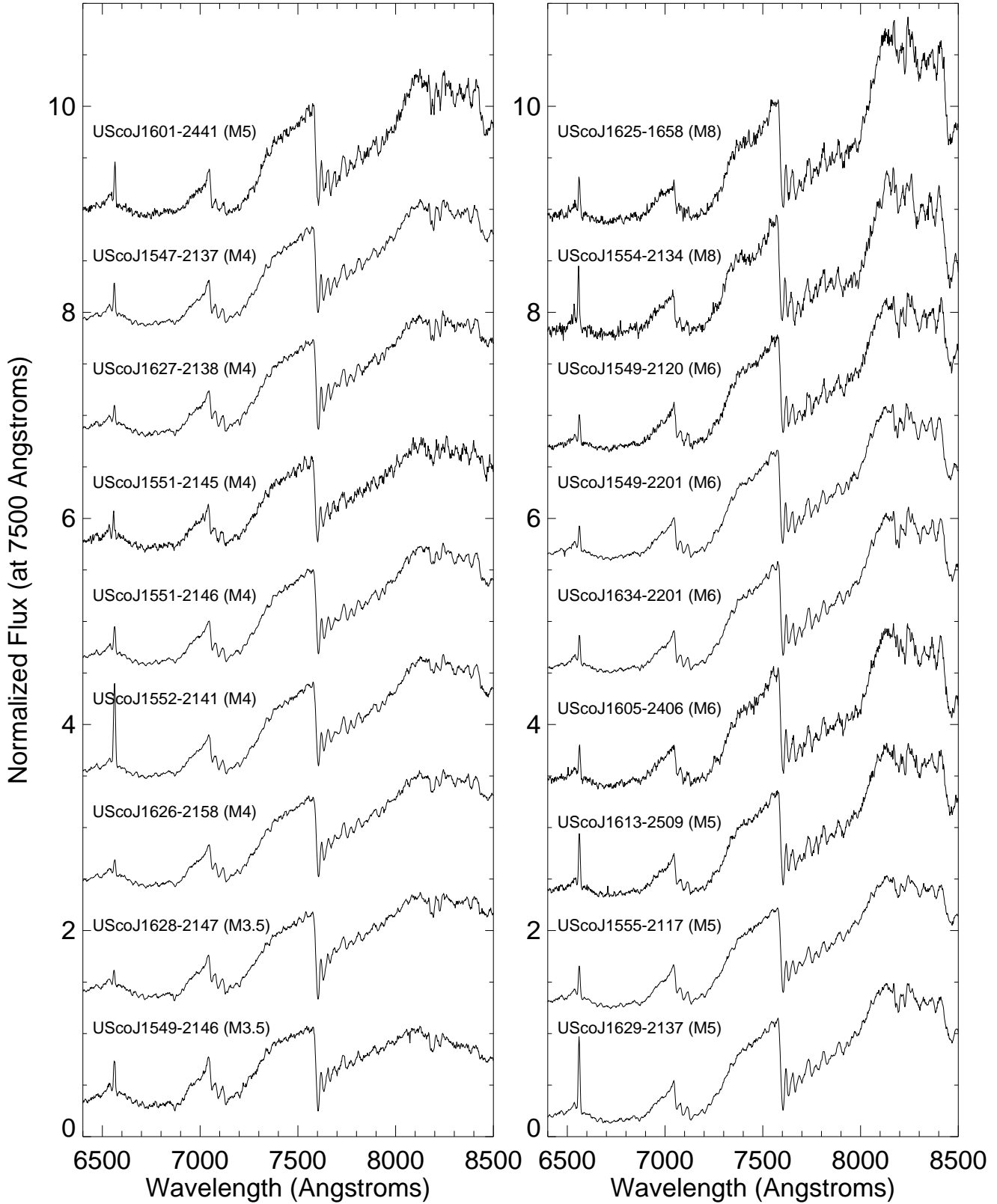}
   \caption{Optical (6400--8500\AA{}) spectra of 15 new 
   members in the Upper Sco association obtained with EFOSC
   on the ESO 3.6-m telescope. The spectral types range from
   M3.5 to M8 with a typical uncertainty of half a subclass. 
   We have added three known members for comparison
   purposes: SCH160147$-$244101 \citep[M5;][]{slesnick06},
   DENIS J160514$-$240653 \citep[M6;][]{martin04}, and
    SCH162528$-$165851 \citep[M8;][]{slesnick06}. All objects show
   strong H$\alpha$ emission and weak Na{\small{I}} doublets,
   indicator of low gravity.
   }
   \label{fig_USco:spectra}
\end{figure*}

The spectral classification of all candidates followed-up 
spectroscopically revealed 3 subgroups (we do not have
spectra for the two faintest candidates):
\begin{enumerate}
\item 12 proper motion non-members (star symbols in Fig.\ \ref{fig_USco:cmds})
classified photometrically as reddened sources are indeed
spectroscopic non-members. All but two sources are reddened 
early-type stars. They are clearly concentrated in one part of 
the association (Fig.\ \ref{fig_USco:coverage})
associated with reddening and possibly due to the
presence of a nearby early-type star. The presence of interstellar
absorption is seen in the IRAS 100$\mu$m map of the association
displayed in Fig.\ 1 in \citet{slesnick06}.
The other two sources are classified as a M5 dwarf
(15$^{h}$50$^{m}$39.55$^{s}$, $-$21$^{\circ}$39$'$47.5$''$)
and the remaining one is a late-M dwarf
(15$^{h}$50$^{m}$11.5$^{s}$, $-$22$^{\circ}$01$'$21.9$''$).
The latter is redder than the coolest Upper Sco member observed
in this study (spectral type of M8) and exhibits H$\alpha$ in
emission. The Na{\small{I}} equivalent width (7.3\AA{}) is
larger than measurements for Upper Sco members and its proper 
motion is inconsistent with the association.
It is also redder than vB10 \citep[M8;][]{kirkpatrick91}
which we included in our program as a field M dwarf template.
We tentatively classify this source as a M9 dwarf.
\item 3 proper motion members classified as spectroscopic non-members
(triangles in Fig.\ \ref{fig_USco:cmds}): two turned out to be
early-type stars, and the remaining one is a M5 dwarf.
\item 15 proper motion members confirmed spectroscopically
as young M dwarfs belonging to the Upper Sco association.
We can subdivide those members into five groups 
(Table \ref{tab_USco:Members}; Fig.\ \ref{fig_USco:spectra}):
2 objects are classified as M3.5 dwarfs, 6 as M4.0,
5 as M5, 3 as M6, and the faintest object is a M8 dwarf.
We assign an uncertainty of half a subclass to our spectral 
types (note that we rely on previous classifications).
Young cluster members with spectral types of M6 or later
are considered as brown dwarfs \citep{martin96,luhman98}.
All but one spectroscopic members lie within
a 2$\sigma$ circle centred on ($-$11,$-$25) mas/yr,
assuming 10 mas/yr as typical errors on the proper
motion measurement.
\end{enumerate}
Consequently, 15 out of 18 proper motion members were confirmed
as spectroscopic members, yielding a contamination
of 3/18 = 16.5\% among our sample, lower than studies
in the Pleiades \citep[31\%;][]{moraux01}
and $\alpha$ Per \citep[30--45\%;][]{barrado02a} open clusters.

%
%
%

%
%
%
\begin{figure}
   \centering
   \includegraphics[width=1.0\linewidth]{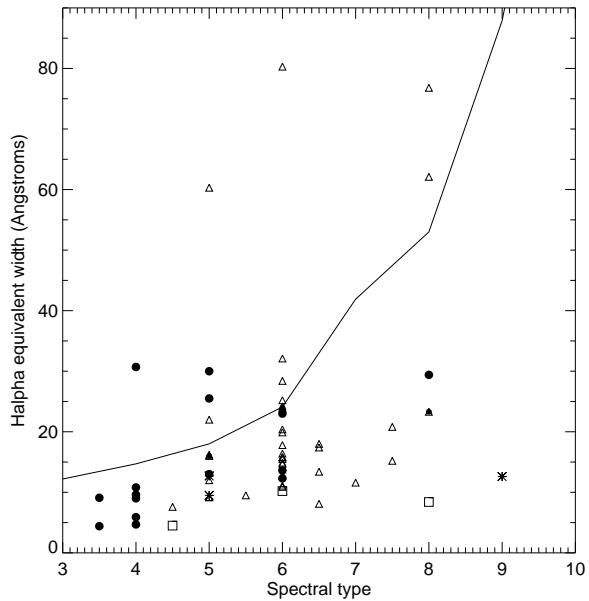}
   \caption{H$\alpha$ equivalent widths as a function of
   spectral type for 15 new spectroscopic members (filled circles)
   and non-members (asterisks).
   The level of chromospheric activity confirms that those 
   candidates as members of the Upper Sco association.
   We have also included equivalent widths for members
   (open triangles) published by \citet{slesnick06} as well
   as for field dwarfs observed with EFOSC2 and show H$\alpha$
   emission (squares).
   Overplotted is the accretor/non-accretor empirical function
   (solid line) defined by \citet{barrado03b}.
   Three objects with spectral types of M4 and M5 lie above
   the solid line and could be accretors.
   }
   \label{fig_USco:Ha_SpT}
\end{figure}
%

%
%
\subsection{Chromospheric activity and gravity}
\label{USco:spectro_Memb}

To assess the youth of the M dwarfs having photometry and proper
motion consistent with Upper Sco, we investigated the
pseudo-equivalent widths of the H$\alpha$ emission line 
(6365\AA{}) and the Na{\small{I}} doublet (8183/8195\AA{}). 

The H$\alpha$ equivalent widths measured for all Upper Sco
members range from $-$4.4\AA{} to 
$-$30.7\AA{} (filled circles in Fig.\ \ref{fig_USco:Ha_SpT})
and lie below the empirical boundary between accreting and 
non accreting low-mass and brown dwarfs defined by 
\citet{barrado03b}. Three objects with strong
H$\alpha$ emission lines lie above this line and are 
discussed in Sect.\ \ref{USco:spectro_activity}.
Finally, we confirm the increasing strength of the
H$\alpha$ equivalent width with later spectral type.

\begin{figure}
   \centering
   \includegraphics[width=1.0\linewidth]{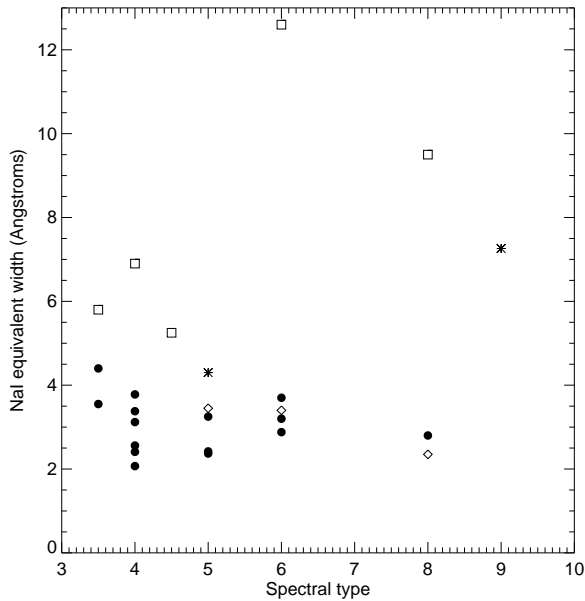}
   \caption{Na{\small{I}} equivalent widths as a function of
   spectral type for 15 spectroscopic new members (filled circles)
   and non-members (asterisks). Three known members 
   published by \citet{martin04} 
   and \citet{slesnick06} were chosen as spectroscopic templates
   and are plotted as diamonds.
   Equivalent widths for 5 field dwarfs targeted with the same 
   telescope and instrument configurations as the Upper Sco 
   sample are displayed as squares.
   The weak Na{\small{I}} doublet equivalent widths measured 
   for the Upper Sco candidate members indicate low gravity and 
   adds credance to their membership.
   }
   \label{fig_USco:NaI_SpT}
\end{figure}

The Na{\small{I}} doublet is a good gravity indicator for spectral
types later than about M3 as it weakens with lower gravity i.e.\ 
younger ages. The equivalent widths measured for the Upper Sco
members (filled circles) and field dwarfs (squares) observed with the
same instrument are plotted as a function of spectral
type in Fig.\ \ref{fig_USco:NaI_SpT}.
Our measurements indicate that young Upper Sco members
exhibit weaker Na{\small{I}} doublets than field dwarfs of similar
spectral type. To put our work into a wider pictures, our
equivalent width measurements tend to be weaker than values
reported in the literature for the Pleiades \citep{martin96}
although the spectral resolution was different. This argument
remain however qualitative as lower resolution tend to
overestimate equivalent widths.

To summarise, 15 out of 18 photometric and proper motion 
selected members have confirmatory signatures of weak 
Na{\small{I}} and H${\alpha}$ emission (see optical spectra 
in Fig.\ \ref{fig_USco:spectra}). Our measurements are in
agreement with results obtained for three known Upper Sco
members published by \citet{martin04} and \citet{slesnick06}
that we observed as templates during our programme.

%
%
%

%
%
%
\begin{figure}
   \centering
   \includegraphics[width=1.0\linewidth]{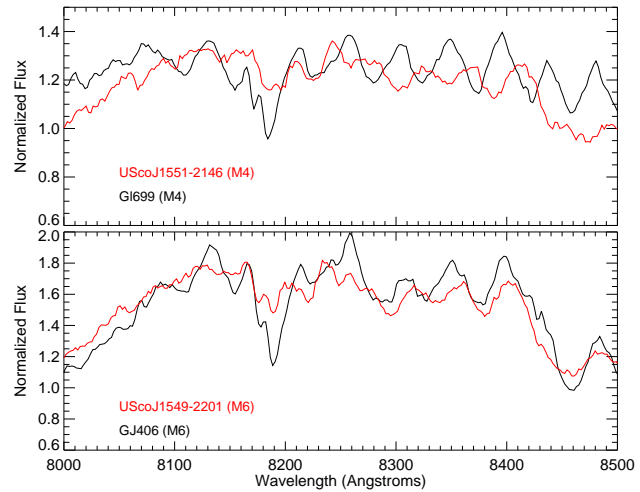}
   \caption{Zoom on the Na{\small{I}} doublet at 
            8183/8195\AA{} for field dwarfs (black) and Upper
            Sco candidates (red) of same spectral type:
            M6 (bottom panel) and M4 (top panel).
            The Na{\small{I}} doublet is gravity sensitive
            i.e.\ the younger the object, the waeker the
            equivalent width is.
   }
   \label{fig_USco:NaI_compare}
\end{figure}

%
%
\subsection{Activity and accretion}
\label{USco:spectro_activity}

Two different mechanisms can account for the H$\alpha$ emission
line detected in all Upper Sco members (Fig.\ \ref{fig_USco:Ha_SpT}):
chromospheric activity is characterised by weak lines with
gaussian profiles whereas accretion and winds present
as strong emission with assymmetric line shapes
\citep{mohanty05}.

We have measured H$\alpha$ equivalent widths which are discrepant
from values reported in the literature for spectroscopic 
templates. We attribute this variation to chromospheric 
activity. We should nevertheless emphasise that the difference in
spectral resolution between our observations and those published
in the literature could account for some of the discrepancy.
The first object,
DENIS J160514$-$240653 \citep{martin04}, was included in our
program as a known Upper Sco member. We measured an
equivalent width of $-$15.7\AA{} in our spectrum compared
to $-$24\AA{} reported in \citet{martin04}
The second object is SCH160311$-$13454, classified as
a M5 field dwarf by \citet{slesnick06}. We have measured
an H$\alpha$ equivalent width of $-$18.8\AA{},
stronger than the one reported previously ($-$8.7\AA{}).
This type of variation in the H$\alpha$ is common among
mid-M field dwarfs \citep{hawley96}.
We have also investigated the Na{\small{I}} doublet for this
object and measured equivalent widths of 5.5\AA{} and 7.0\AA{}
in our spectrum and Slesnick's spectrum, respectively.
Those values are suggestive of an age older than Upper Sco.
We also measured a proper
motion of ($+$90,$-$54) mas/yr between the 2MASS observations
and the optical survey by \citet{slesnick06}, clearly
inconsistent with the association.

The empirical relation defined by \citet{barrado03b} from
low-resolution optical spectroscopy to distinguish
accreting and non-accreting low-mass stars and brown dwarfs
is overplotted in Fig.\ \ref{fig_USco:Ha_SpT}.
Most members lie below this empirical relation. However,
three objects, one M4 and two M5 dwarf members of Upper Sco
exhibit H$\alpha$ equivalent widths two to three times those of 
sources with the same spectral type and larger than the
divide line between accretors and non-accretors.
Although weaker than the strongest H$\alpha$ equivalent 
widths reported by \citet{slesnick06} in several
objects (open triangles in Fig.\ \ref{fig_USco:Ha_SpT}) and
despite our low spectral resolution which might have
overestimate the equivalent widths,
those sources could still undergo accretion
unless we observed strong flares due to chromospheric activity.
Higher resolution spectroscopy centred on H$\alpha$
is required to investigate the line profiles and the
origin of the strong emission.

%
%
\subsection{Binarity}
\label{USco:binary}
\citet{bouy06b} reported a possible excess of wide binaries 
with separations between 100 and 150 AU among low-mass stars 
and brown dwarfs in Upper Sco.
Similarly, \citet{luhman05} discovered serendipitously a wide 
low-mass pair with a projected physical separation of 130 AU
in the association.

A couple objects brighter than $J$ = 14 mag lie clearly to the
right of the cluster sequence which runs almost vertically 
in the ($J-K$,$J$) colour-magnitude diagram, 
suggesting that they are possible low-mass binaries. 
The object at $J$ = 15.5 mag (M $\sim$ 0.015 M$_{\odot}$) 
could also be a multiple substellar system and of particular 
interest to test evolutionary models at low masses 
(if confirmed as such). 
Other binary candidates might be hidden in the list of potential
reddened members.
Because we selected only point sources in the EDR catalogue,
we are not sensitive to separations of about 1-2 pixel or
0.4--0.8 arcsec. This corresponds to projected physical
separations of 50--100 AU at the distance of the association
(d = 145 pc). We are also insensitive to closer companions 
as it requires high-resolution imaging or adaptive optics
although the location of objects in the various colour-magnitude
diagrams can give some clues about the possible multiplicity
of a source.
Finally, we should be sensitive to larger separations and
be able to detect wide companions on seeing-limited images
(separation larger than a few hundreds of astronomical units)
especially on the $K$-band images where microstepping is used
to measure accurate proper motions when the second epoch 
observations are released.

%
%
\section{Summary}
\label{USco:summary}

We have presented the analysis of a 9.3 square degree survey in the
Upper Scorpius association conducted for the UKIDSS Galactic
Cluster Survey and extracted from the Early Data Release.
Combining infrared photometry, proper motion, and optical
spectroscopy, the membership of the original 112 candidates
can be summarised as follows:
\begin{enumerate}
\item 80 photometric candidates selected from the ($Z-J$,$Z$)
colour-magnitude diagram were rejected on the basis their $J-K$
affected by reddening. Among them, 15 sources could be reddened
members as they exhibit proper motion consistent with the
association
\item 13 photometric candidates have proper motions inconsistent
with the Upper Sco association and optical spectra of early-type
stars affected by reddening
\item 3 objects have optical spectroscopy inconsistent with 
membership despite their photometry and proper motion suggesting 
that they are possible members
\item 15 candidates are definite members confirmed 
from photometry, proper motion, and optical spectroscopy, 
including 4 new brown dwarfs (spectral types later than M6).
They span masses between 0.2 and 0.01 M$_{\odot}$, assuming an
age of 5 Myr and a distance of 145 pc for Upper Sco
\item 2 faint photometric candidates with estimated masses below
20 M$_{\rm Jup}$ according to the DUSTY models \citep{chabrier00c}.
One object has proper motion consistent with Upper Sco and no
spectroscopy whereas the other one has neither proper motion
nor spectroscopy currently available
\end{enumerate}
We detected variation in the H$\alpha$ equivalent widths 
for one Upper Sco members and one field dwarf, hence showing
signs of chromospheric activity. In addition, 3 new
members exhibit H$\alpha$ emission lines stronger than the 
lower limit of the empirical relation between accreting 
and non accreting low-mass stars and brown dwarfs.
Finally, a handful of members present a high probability 
of being binary systems from their location in the colour-magnitude 
diagrams.

%
%
\section*{Acknowledgments}

NL is a postdoctoral research associate funded by PPARC UK.
We are grateful to Isabelle Baraffe and France Allard for providing 
us with the NextGen and DUSTY models for the WFCAM filters
and Eduardo Mart{\'{\i}}n and Catherine Slesnick for kindly
supplying their optical spectra.
We thank our colleagues at the UK Astronomy Technology Centre,
the Joint Astronomy Centre in Hawaii, the Cambridge Astronomical Survey
and Edinburgh Wide Field Astronomy Units for building and operating
WFCAM and its associated data flow system.
This research has made use of the Simbad database, operated at
the Centre de Donn\'ees Astronomiques de Strasbourg (CDS), and
of NASA's Astrophysics Data System Bibliographic Services (ADS).
This publication makes use of data products from the
Two Micron All Sky Survey, which is a joint project of the
University of Massachusetts and the Infrared Processing and
Analysis Center/California Institute of Technology, funded by
the National Aeronautics and Space Administration and the National
Science Foundation.

%
%
\bibliographystyle{mn2e}
\bibliography{../../AA/mnemonic,../../AA/biblio_old}



\appendix

\section{SQL query submitted to the WFCAM Science Archive}
\label{sql}

Initial sample selection for this Upper Sco study was made by accessing
the UKIDSS Early Data Release database (UKIDSSEDR) held at the WFCAM 
Science Archive. The SQL query given in Figure~\ref{sqlquery} was used.

\begin{figure*}
\begin{verbatim}
SELECT 
/*    Attribute selection:                                               */
      g.ra, g.dec, zmyPnt, ymjPnt, jmhPnt, hmk_1Pnt,
      zaperMag3, yaperMag3, japerMag3, haperMag3, k_1aperMag3, 
      3.6e6*COS(RADIANS(g.dec))*(g.ra-T2.ra)/((mj.mjdObs - T2.jdate+2400000.5)/365.25) AS pmRA, 
      3.6e6*(g.dec-T2.dec)/((mj.mjdObs - T2.jdate+2400000.5)/365.25) AS pmDEC

FROM 
/*    Table(s) from which to select the attributes:                      */
      gcsMergeLog AS l, Multiframe AS mj, (
         SELECT t.ra AS ra, t.dec AS dec, x.slaveObjID AS slaveObjID,
                x.masterObjID as masterObjID, t.j_m, t.h_m, t.k_m, t.jdate
         FROM   gcsSourceXtwomass_psc as x, TWOMASS..twomass_psc as t
         WHERE  x.slaveObjID=t.pts_key AND distanceMins IN (
            SELECT MIN(distanceMins) FROM gcsSourceXtwomass_psc WHERE  masterObjID=x.masterObjID
         )
      ) AS T2 RIGHT OUTER JOIN gcsSource AS g ON (g.sourceID=T2.masterObjID)

WHERE 
/*    Sample selection predicates:                             
      only Upper Sco data (no other GCS target is south of the equator)  */
      g.dec < 0.0
/*    Bright saturation cut-offs                                         */
      AND zaperMag3 > 11.4
      AND yaperMag3 > 11.3 
      AND japerMag3 > 10.5 
      AND haperMag3 > 10.2 
      AND k_1aperMag3 > 9.7
/*    Limit merged passband selection to +/- 1 arcsec                    */
      AND zXi BETWEEN -1.0 AND +1.0
      AND yXi BETWEEN -1.0 AND +1.0
      AND jXi BETWEEN -1.0 AND +1.0
      AND hXi BETWEEN -1.0 AND +1.0
      AND k_1Xi BETWEEN -1.0 AND +1.0
      AND zEta BETWEEN -1.0 AND +1.0
      AND yEta BETWEEN -1.0 AND +1.0
      AND jEta BETWEEN -1.0 AND +1.0
      AND hEta BETWEEN -1.0 AND +1.0
      AND k_1Eta BETWEEN -1.0 AND +1.0
/*    Retain only point-like sources                                     */
      AND zClass BETWEEN -2 AND -1
      AND yClass BETWEEN -2 AND -1
      AND jClass BETWEEN -2 AND -1
      AND hClass BETWEEN -2 AND -1
      AND k_1Class BETWEEN -2 AND -1
/*    Retain only the best record when duplicated in an overlap region   */
      AND (priOrSec = 0 OR priOrSec = g.frameSetID)
/*    Table join predicates:                                             */
      AND g.frameSetID=l.frameSetID
      AND l.jmfID=mj.multiframeID
\end{verbatim}
\caption[]{Structured Query Language (SQL) query used on the WFCAM Science
Archive database UKIDSSEDR to select the GCS Upper Sco sample discussed in
the paper. The query returns 174,010 rows of data. }
\label{sqlquery}
\end{figure*}

For more details concerning the use of SQL and data in the 
WFCAM Science Archive, see Hambly et al.~(2006) and references
therein.

\vfill

\bsp

\label{lastpage}

\end{document}